\documentclass[aps,prb,twocolumn,showpacs,10pt]{revtex4-1}
\usepackage{graphicx,amsmath,amssymb,amsfonts,sidecap,color}

\usepackage{color}
\usepackage[usenames,dvipsnames,svgnames,table]{xcolor}

\newcommand{\e}[1]{{\rm e}^{#1}}

\makeatletter
\newcommand*{\rom}[1]{\expandafter\@slowromancap\romannumeral #1@}
\makeatother

\begin{document}

\title{Fermionic and Majorana Bound States in  Hybrid Nanowires \\ with Non-Uniform Spin-Orbit Interaction }

\author{Jelena Klinovaja$^{1}$}
\author{Daniel Loss$^{2,3}$}
\affiliation{$^1$Department of Physics, Harvard University,  Cambridge, Massachusetts 02138, USA,}
\affiliation{$^2$Department of Physics, University of Basel,
             Klingelbergstrasse 82, CH-4056 Basel, Switzerland}
\affiliation{ $^3$RIKEN Center for Emergent Matter Science, 2-1 Hirosawa, Wako, Saitama, 351-0198 Japan}

\date{\today}
\pacs{74.45.+c; 73.21.Hb}

\begin{abstract}
We study intragap bound states in the topological phase of a  Rashba nanowire in the presence of a magnetic field and with non-uniform spin orbit interaction (SOI) and proximity-induced superconductivity gap.
We show that fermionic bound states (FBS) can emerge inside the  proximity gap. They are localized at the junction between two wire sections characterized by different directions of the SOI vectors, and 
they coexist with Majorana bound states (MBS) localized at the nanowire ends.
The energy of the FBS is determined by the angle between the SOI vectors and the lengthscale over which the SOI changes compared to the Fermi wavelength and the localization length. 
We  also consider double-junctions 
and show that the two emerging FBSs can hybridize and form
a double quantum dot-like structure inside the gap. We find explicit analytical solutions of the bound states and their energies for certain parameter regimes such as weak and strong SOI. The analytical results are confirmed and complemented by an
independent numerical tight-binding model approach. 
Such FBS can act as quasiparticle traps and thus can have implications for topological quantum computing schemes based on braiding MBSs.
\end{abstract}

\maketitle

\section{Introduction}

Bound states arising in a variety of condensed matter systems were explored intensively over the last decade or so. Some of these states such 
as Majorana bound states (MBSs),\cite{fu,Nagaosa_2009,Sato,lutchyn_majorana_wire_2010,oreg_majorana_wire_2010,alicea_majoranas_2010,MF_ee_Suhas,potter_majoranas_2011,Klinovaja_CNT,bilayer_MF_2012,Rotating_field,
RKKY_Basel,RKKY_Franz,RKKY_Simon,MF_nanoribbon,MF_MOS,Shiba_ladder,Ali,MF_Bena,MF_dot,Bena_MF,Pascal,mourik_signatures_2012,deng_observation_2012,das_evidence_2012,Rokhinson,Goldhaber,marcus_MF} fractional fermions,
\cite{JR_model,CDW_suhas,FF_1,SSH_model,FF_non_Abelian,FF_transport,FF_pump}
and parafermions\cite{Fradkin_PF_1980,topology_barkeshli,barkeshli_2, Fendley_PF_2012,PF_Linder,Cheng,Vaezi,PF_Clarke,
Ady_FMF,PF_Mong,vaezi_2,PFs_Loss,PFs_Loss_2,PF_TI,PF_Thomas,PF_Oreg,Vaezi_2} are attractive due to their potential use in topological quantum
computation schemes owing to their non-Abelian statistics.~\cite{Stern_review,Alicea_review} Moreover,  MBSs can be transformed into fractional fermions by tuning parameters from
the topological phase to the trivial phase.\cite{Rotating_field}
At the same time, the possibility of generating MBSs together with fermionic bound states (FBSs) simultaneously, in the topological phase, has not received much attention so far. 

In this work we explore such a scenario in non-uniform Rashba nanowires coupled to an $s$-wave superconductor 
in the presence of a magnetic field in the topological regime ({\it i.e.} with Zeeman energy being dominant over the superconducting  pairing). 
The considered non-uniformity is in the spin orbit interaction (SOI) along the wire. Specifically,
we consider a situation where the SOI vector changes its direction along the nanowire axis creating an interface between two nanowire sections with different SOI vector directions. 
In this case, we  
find that additional FBSs localized at 
such SOI interfaces emerge. In contrast to zero-energy MBSs localized at the nanowire ends, the energy of these FBSs
crucially depends on the SOI vector rotation angle and can take any value inside the proximity induced bulk gap.
In the particular case, when 
the SOI has a sharp discontinuity, {\it i.e.}, changes abruptly,  such that
the SOI vector rapidly rotates by the angle $\phi=\pi$, or, equivalently, changes its sign, the system possesses an additional symmetry that constrains the FBSs to be zero-energy states.\cite{pi_junction} However, this degeneracy
between the FBS being filled and unfilled
is not robust against small perturbations. For instance, if the rotation angle deviates from $\pi$, the FBS energy levels smoothly shifts away from zero and eventually disappears from the gap into the continuum at $\phi=0$. The localization length of such states is the larger the closer the energy level is to the gap edge.

In addition, we also study the situation of the SOI vector rotation taking place over a finite region $\ell$ of the nanowire, which could be comparable to the Fermi wavelength $\lambda_F$ and the FBSs localization length $\xi$. As expected, if the SOI vector rotates adiabatically, $\ell>\lambda_F,\xi$, the FBS energy level merges into the continuum, such that no FBSs emerge at the interface.
We emphasize that the FBSs exist at the interface (or junction) between two sections where both of them are in the topological phase.
This is in contrast to MBSs, which exist at the interface between the topological and non-topological phase.

Besides fundamental interest, the study of non-uniform SOI
is also important for practical implementations of topological quantum computing schemes based on MBSs.~\cite{Alicea_review} 
The non-uniformity of the SOI could be a result
of a change in the direction of an electric field causing the Rashba SOI~\cite{Rashba_1960,SOI_book} or of a change in the direction of the crystallographic axes in systems with 
Dresselhaus SOI.~\cite{Dresselhaus_1958,SOI_book} Such changes are likely to occur, for example, in  $T$-junctions that underlay proposed schemes for braiding of MBSs.~\cite{Alicea_Fisher} 

Moreover, using the gauge equivalence in one-dimensional systems between Rashba SOI plus uniform magnetic field and a spatially rotating magnetic field but without SOI,\cite{Braunecker}
one can easily show that a spatial discontinuity in the SOI in a hybrid Rashba nanowire is equivalent to a {\it domain wall} in a hybrid RKKY system with self-tuned topological phase.\cite{RKKY_Basel,RKKY_Simon,RKKY_Franz}  Thus, here again, we can  expect FBSs to emerge simultaneously with MBSs.
If the RKKY system, in addition, has an easy plane
 the domain walls are $\pi$-junctions, such that FBSs can efficiently hybridize with MBSs.

This then raises the  question whether such additional bound states could affect the properties of topological braiding schemes based on MBSs or affect the decoherence of MBSs themselves due to trapping and releasing of fermions in such FBSs.
Here, we will not address this important issue further but instead will focus on the physics of such FBSs and on the conditions under which they exist and what their behavior is as function of various system parameters.

The paper is organized as follows. In Sec. \ref{sec:model} we describe the system  under  consideration and provide the effective Hamiltonian both for the analytical and
numerical models.  We begin with addressing the presence of FBSs in the regimes of strong (see Sec. \ref{sec:strong}) and weak SOI (see Sec. \ref{sec:weak}) under the assumption
of an abrupt change in the SOI vector direction. The double SOI-junctions are considered in Sec. \ref{sec:junction}.
Subsequently, in Sec. \ref{sec:smooth}, we extend our results also to the case of smoothly varying SOI interaction. 
The final section, Sec. \ref{sec:conc}, contains our conclusions.

\section{Model\label{sec:model}}

\subsection{Analytical Model}

\begin{figure}[!b]
\includegraphics[width=0.8\linewidth]{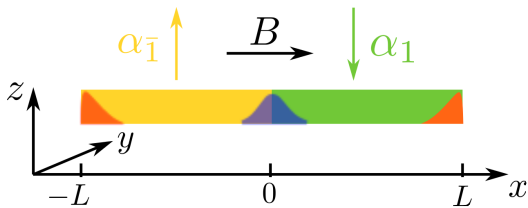}
\caption{Sketch of a nonuniform nanowire of length $2L$ directed along $x$ direction which consists of two segments $x<0$ (left section, yellow) and $x>0$ (right  section, green) in which the corresponding SOI vectors ${\boldsymbol  \alpha}_{\bar 1}$ and ${\boldsymbol  \alpha}_1$ point in different directions in the $yz$ plane.
If the nanowire is in the topological regime such that the Zeeman energy produced by a magnetic field $\bf B$ applied along the nanowire dominates over the proximity-induced superconductivity gap,
bound states are formed inside the gap. First, there are zero-energy Majorana bound states (red area) localized at the nanowire ends $x=\pm L$. Second, there are fermionic bound states (blue area) at the interface $x=0$, where the SOI vector changes its direction. These fermionic bound states are not fixed to zero energy and can acquire any energy inside the gap depending on the relative angle between the two SOI vectors ${\boldsymbol  \alpha}_{\bar 1}$ and ${\boldsymbol  \alpha}_1$.
} 
\label{fig:model}
\end{figure}

We consider a nanowire with Rashba SOI  brought into proximity to an $s$-wave superconductor in the presence of an applied magnetic field $\bf B$, see Fig.~\ref{fig:model}. The SOI in the nanowire is characterized by the SOI vector $\boldsymbol \alpha$. The direction of
$\boldsymbol \alpha$ determines the direction in which the spins are polarized by the SOI, and its magnitude $\alpha$  determines the SOI strength. A particularly interesting regime emerges when the direction of  $\bf B$ is perpendicular to $\boldsymbol \alpha$. If the resulting Zeeman energy  dominates over the proximity-induced superconductivity gap in the nanowire, the system hosts MBSs localized at the  ends of the nanowire.~\cite{lutchyn_majorana_wire_2010,oreg_majorana_wire_2010,alicea_majoranas_2010} Below we extend this well-known uniform model to the case of non-uniform SOI, namely, to  a setup in which the SOI vector $\boldsymbol \alpha (x)$ changes its direction as a function of position along the nanowire. In what follows, without loss of generality, we fix the direction of the SOI vector $\boldsymbol \alpha$
to be in the $z$ direction for $x<0$, while
for $x>0$  
it lies in the $yz$ plane,
\begin{align}
\boldsymbol \alpha (x) = \begin{cases}
\alpha_{\bar 1} \hat {\bf z}, &x<0\\
\alpha_{1} (\bf \hat z \cos \phi \  +\bf \hat y \sin  \phi \ ), &x>0
\end{cases}.
\end{align}

The kinetic part of the Hamiltonian is given by
\begin{align}
H_{0}= \sum_{\sigma=\pm1} \int dx\ \Psi_{\sigma}^\dagger (x) \left[-\frac{\hbar^2\partial_x^2}{2m} -\mu (x)\right]\Psi_{\sigma} (x),
\end{align}
where the fermionic annihilation operator $\Psi_{\sigma} (x)$ removes an electron (of charge $e$ and effective mass $m$)  with spin $\sigma=\pm 1$  at the position $x$. The integral over $x$ runs over the entire wire of length $2L$.
The spin quantization axis is chosen to be along the $z$ direction.  The chemical potential $\mu (x)$ is assumed to be  uniform in each of the two nanowire segments, but changes between the segments, where  $\mu(x)\equiv \mu_j$ is tuned to the corresponding SOI energy $E_{SOI,j}= \hbar^2 k_{so,j}^2/2m$ with the SOI wavevector $k_{so,j} =  m \alpha_j/\hbar^2$. Here, $j= -1, 1,$ distinguishes the left ($-L<x<0$) and the right ($0<x<L$) segment of the nanowire, respectively.

The SOI term is written as
\begin{align}
&H_{SOI} = -\frac{i}{2}  \sum_{\sigma,\sigma'=\pm 1}\int dx\ \nonumber\\
&\hspace{10pt}\Psi_{\sigma}^\dagger (x) ([\boldsymbol \alpha (x)  \cdot \boldsymbol \sigma]_{\sigma\sigma'} \partial_x +\partial_x [\boldsymbol \alpha (x) \cdot \boldsymbol \sigma]_{\sigma\sigma'} ) \Psi_{\sigma'} (x),
\label{SOI_2}
\end{align}
where $\sigma_k$ are Pauli matrices with $k=x,y,z$. Here, we use the symmetrized form of the SOI such that $H_{SOI}$ is a hermitian operator. 
Any change of $\boldsymbol \alpha (x)$ as function of position $x$ we shall consider in the following is assumed to be smooth on the atomistic scale. This is for self-consistency reasons, {\it i.e.},  to 
preserve  the validity of the effective Rashba or Dresselhaus SOI derived from the bandstructure in the low-energy and long-wavelength limit.

The magnetic field $\bf B$ applied along the $x$-axis such that it always stays perpendicular to $ \boldsymbol \alpha (x)$ results in a term
\begin{equation}
H_Z=\int dx\ \Delta_Z \Psi_{\sigma}^\dagger (x) (\sigma_x)_{\sigma\sigma'}  \Psi_{\sigma'} (x),
\end{equation}
where the Zeeman energy is given by $\Delta_Z=g\mu_B B/2$, with $g$ being the $g$-factor and $\mu_B$  the Bohr magneton.

If the nanowire is placed in contact with an $s$-wave superconductor, the proximity induced superconductivity of  strength  $\Delta_{sc}$ is described by the term
\begin{align}
&H_{SC} =\frac{\Delta_{sc }}{2}\int dx\ \Big[ \Psi_{1}^\dagger \Psi_{\bar 1}^\dagger  - \Psi_{\bar 1}^\dagger \Psi_{ 1}^\dagger + H.c.\Big].
\label{SC}
\end{align}
The total Hamiltonian under  consideration is the sum of all terms mentioned above and given by
\begin{align}
H=H_0+H_{SOI}+H_Z+H_{SC}.
\label{sum}
\end{align}
The bulk energy spectrum is then readily obtained,
\begin{align}
&E_{\pm,j}^2= \left(\frac{\hbar^2 k^2}{2m}\right)^2 + (\alpha_i k)^2 + \Delta_Z^2 + \Delta_{sc}^2\nonumber\\
&\hspace{40pt}\pm  2 \sqrt{\Delta_Z^2 \Delta_{sc}^2 + \left(\frac{\hbar^2 k^2}{2m}\right)^2 [(\alpha_j k)^2 + \Delta_Z^2]},
\end{align}
where $j= \pm 1$ corresponds to the right/left segment of the nanowire, see Fig.~\ref{fig:model}. Importantly, independent of the SOI strength and of the SOI direction,  the bulk spectrum is  gapless if $\Delta_Z=\Delta_{sc}$ for both nanowire segments. 
This is a signature of the topological phase transition that separates the topological phase with zero-energy MBSs localized at the nanowire ends from the trivial phase without such bound states.~\cite{lutchyn_majorana_wire_2010,oreg_majorana_wire_2010,alicea_majoranas_2010} At the same time, the system in the trivial phase can also host fermionic bound states with their  energy also lying inside the bulk gap.\cite{Rotating_field} However, in contrast to MBSs, their energy is sensitive to  local fluctuations in the system parameters such as, for example, the chemical potential.\cite{FF_non_Abelian}
In this work, we explore if it is possible to generate FBSs simultaneously with MBSs, {\it i.e.} in the topological phase. As we show below, the answer is positive for non-uniform nanowires in which the SOI vector changes its direction. In this case, the FBSs are localized in the region of the nanowire where the SOI vector $\boldsymbol \alpha (x)$ rotates in space.

\subsection{Numerical Model}

The Hamiltonian $H$ introduced in the previous subsection [see Eq. (\ref{sum})] can be modeled in the tight-binding framework as
\begin{align}
&H_0^n = \sum_{i=1}^N \sum_{\sigma =\pm 1} [-t_x (c_{(i+1)\sigma}^\dagger c_{i\sigma} + c_{i\sigma}^\dagger c_{(i+1)\sigma}) +\mu c_{i\sigma}^\dagger c_{i\sigma}], \label{tb2} \\
&H_{SOI}^n=- \frac{i}{2} \sum_{i=1}^{N-1} \sum_{\sigma,\sigma' =\pm 1} c_{(i+1) \sigma}^\dagger [(\boldsymbol{\bar \alpha}_{i+1} +\boldsymbol{\bar \alpha}_i)\cdot \boldsymbol \sigma]_{\sigma \sigma'} c_{i \sigma'}\nonumber\\
&\hspace{150pt}+ H.c.,\\
& H_Z^n= \sum_{i=1}^N \sum_{\sigma =\pm 1} \Delta_Z c_{i\sigma}^\dagger c_{i\bar\sigma},\\
&H_{sc}^n= \sum_{i=1}^N \Delta_{sc} (c_{i1}^\dagger c_{i\bar1}^\dagger + c_{i\bar1} c_{i1}).
\label{tb1}
\end{align}
Here, the hopping amplitude $t_x = \hbar^2/(2ma^2)$ sets the width of the band, and $a$  is the lattice constant. 
[Parenthetically, we note that the lattice constant $a$ of the tight-binding model is typically ten times smaller than the Fermi wavelength but, for self-consistency, is implicitly assumed to be much larger than the true atomistic lengthscale of the modeled semiconductor.]
 The number of sites $N$ in the chain  sets the nanowire length $2L\equiv(N-1)a$. The SOI vector $\boldsymbol{\bar \alpha}_i $ determines the SOI energy $E_{SOI} =\bar \alpha_i^2/t_x$ and is connected to the SOI vector from the previous section via $\boldsymbol{\bar \alpha}_i  = \boldsymbol{\alpha}|_ {x=(i-N/2)a}/(2a)$. In what follows, we again consider the SOI vector  to lie in the $yz$-plane,
\begin{align}
\boldsymbol{\bar \alpha}_i = {\bar \alpha}({ \bf\hat z} \cos \phi_i + {\bf\hat y} \sin \phi_i ).
\end{align}
The angle $\phi_i$ varies from $\phi_1=0$ at the left  end to $\phi_N=\phi$ at the right end of the nanowire. 
The SOI angle $\phi$ varies as a function of  position as
\begin{align}
\phi_i= \frac{\phi}{2} \{1+{\rm tanh}[(2i-N-1)/l]\}.
\end{align}
Here, $l$ is the characteristic width over which the SOI vector rotates. We distinguish two limiting cases. In the first case, the SOI vector rotates rapidly 
over a distance $l$ much smaller than the Fermi wavelength $\lambda_F=2\pi/k_F$, {\it i.e.}, $l\ll \lambda_F$, whereas in the second one, the SOI vector rotates adiabatically, {\it i.e.}, $l\gg \xi>\lambda_F$. Here $\xi$ is the MBS localization length. Again,  for self-consistency we implicitly assume  that $l$ is larger than any atomistic scale of the described semiconductor.
Based on this model we can find the various bound states numerically and compare with the analytical solutions obtained in some limits as we discuss in the following sections.

\section{Regime of strong Spin Orbit Interaction \label{sec:strong}}

\begin{figure}[!t]
\includegraphics[width=0.75\linewidth]{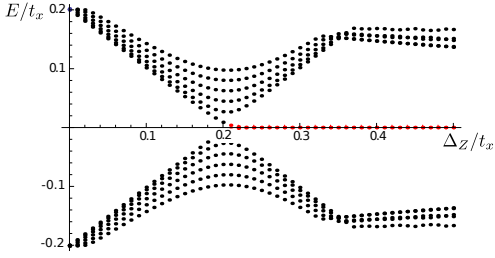}
\caption{The energy spectrum of the nanowire with uniform SOI, $\phi=0$, as function of the Zeeman energy $\Delta_Z$ found in the tight-binding model [see Eqs. (\ref{tb2}) - (\ref{tb1})]. The chain consists of $N=100$ sites. The parameters are chosen as $\bar \alpha/t_x=0.3$, $\Delta_{sc}/t_x=0.2$, and  $\mu$ is tuned to $E_{SOI}$. As expected, if $\Delta_Z=\Delta_{sc}$, the extended bulk states  (black dots) move to zero energy and close the gap. If $\Delta_Z<\Delta_{sc}$, there are no bound states in the gap. However, if $\Delta_Z>\Delta_{sc}$, the system is in the topological phase with two MBS at zero-energy  (red dots). One MBS is localized at the right and  the other one  at the left end of the nanowire. 
}
\label{closing}
\end{figure}

\subsection{Linearized Effective Hamiltonian}
In this section, we focus on the regime of strong spin-orbit interaction where the SOI energy $E_{SOI}$ is much larger than the proximity gap $\Delta_{sc}$ and the Zeeman energy $\Delta_Z$. Here, for a moment, we focus on the left section of the nanowire $x<0$, where the SOI vector is strictly along the $z$-axis. We note that the eigenvectors for the right section $x>0$ can be obtained by rotating the spin basis by an angle $\phi$ around the $x$-axis and by changing the SOI strength.

The Hamiltonian $H$ can be linearized around the Fermi points, which  are determined solely by the SOI and are equal to $k_{F0}=0$ and $\pm k_{F \bar 1}=\pm 2k_{so,\bar 1}$. The electron annihilation operator $\Psi_\sigma(x)$ can be represented around these Fermi points as a sum of slowly-varying right- [$R_\sigma(x)$] and left- [$L_\sigma(x)$] moving fields
\begin{align}
&\Psi_1(x)=R_1 e^{2ik_{so,\bar 1}x} + L_1,\\
&\Psi_{\bar 1}(x)=  R_{\bar 1} + L_{\bar 1} e^{-2ik_{so,\bar 1}x}.
\end{align}

\begin{figure}[!t]
\includegraphics[width=0.7\linewidth]{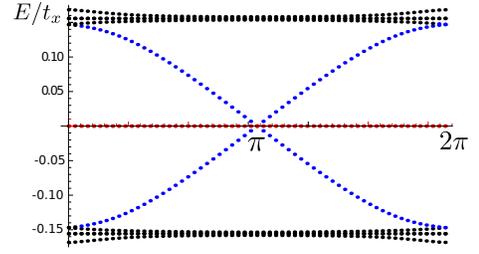}
\caption{The energy spectrum of the nanowire in the topological phase as function of the SOI angle $\phi$ calculated in the tight-binding model [see Eqs. (\ref{tb2}) - (\ref{tb1})]. The chain consists of $N=100$ sites. The parameters are fixed to be $\bar \alpha/t_x=0.3$, $\Delta_{sc}/t_x=0.2$, and $\Delta_Z/t_x=0.4$, and $\mu$ is tuned to $E_{SOI}$.  There is one MBS at the left nanowire end and one MBS at the right nanowire end  (red dots). In addition, there are two FBSs localized at $x=0$ (blue dots) with energy depending on $\phi$. In particular, for opposite SOI vectors, $\phi=\pi$, the FBSs are at zero energy and, thus, degenerate with the MBSs, whereas for $\phi=0,2\pi$, the FBS merge with the extended bulk states (black dots).
}
\label{no_overlap}
\end{figure}

The kinetic part of the Hamiltonian $H_{kin}=H_0+H_{SOI}$  then takes on the following form,
\begin{align}
H_{kin}^s= -i \hbar \upsilon_{F,\bar 1} \sum_{\sigma} \int dx\  (R_\sigma^\dagger \partial_x R_\sigma -L_\sigma^\dagger \partial_x L_\sigma),
\end{align}
where the Fermi velocity $ \upsilon_{F,\bar 1}$  is given by $ \upsilon_{F,\bar 1} = \hbar k_{so,\bar 1}/m$.
The Zeeman term is given by
\begin{align}
H_Z^s = \Delta_Z \int dx\ (R_{\bar 1}^\dagger L_{ 1}+L_1^\dagger R_{\bar 1}),
\end{align}
while the superconductivity term is given by
\begin{align}
H_{SC}^s = \frac{\Delta_{sc}}{2}  \int dx\ (R_1^\dagger L_{\bar 1}^\dagger + L_{\bar 1}^\dagger R_1^\dagger + H.c.).
\end{align}
The Hamiltonian density $\mathcal H^s$ is written in terms of the Pauli matrices $\sigma_i$, $\eta_i$ (acts on electron-hole space), and $\tau_i$  (acts on right-/left-mover space) as
\begin{align}
\mathcal H^s = \hbar \upsilon_ {F,\bar1} \hat k \tau_3 + \Delta_{sc} \sigma_2 \eta_2 + \Delta_Z (\tau_1\sigma_1 - \tau_2 \sigma_2 )\eta_3 /2,
\end{align}
where $\hat k = -i \hbar  \partial_x$ is the momentum operator.\cite{MF_composite}
The corresponding bulk  spectrum is given by
\begin{align}
&E_{e}^{\pm}=\pm\sqrt{(\hbar \upsilon_ {F,\bar1}  k)^2 + \Delta_{sc}^2},\label{bulk_1}\\
&E_{i,\pm}^{\pm}=\pm\sqrt{(\hbar \upsilon_ {F,\bar1} k)^2 + (\Delta_{sc} \pm \Delta_Z)^2},\label{bulk_2}
\end{align}
where the energy levels $E_{e}^{\pm}$ are twofold degenerate. As was already shown above, the system is gapped for all non-zero parameter values except when $\Delta_Z=\Delta_{sc}$, which correspond to the topological phase transition (see Fig. \ref{closing}). Moreover, we note that the bulk spectrum is independent of the direction of the SOI vector in the $yz$-plane. Thus, the bulk spectrum at the right section of the wire, $x>0$, is also given by Eqs.~(\ref{bulk_1})-(\ref{bulk_2}) with the corresponding exchange of the Fermi velocity $\upsilon_ {F,\bar1} \to \upsilon_ {F,1} $.

\subsection{Bound States}

Now we focus on bound states localized around $x=0$, {\it i.e.}, at the interface between two sections of the nanowire with different directions of the SOI vector. This case corresponds to a finite discontinuity in the SOI, {\it i.e.}, to an abrupt change of the SOI at the interface. 
In contrast to MBSs which are always zero-energy  bound states, FBSs can also be at non-zero energy $\epsilon$ inside the bulk gap. The spectrum of the FBSs can be found following standard scattering theory approach. 

First, we identify eigenstates of $\mathcal H^s$ both at the left and right section of the nanowire at a given energy $\epsilon$. In addition, for a moment, we focus on an infinite nanowire and impose the boundary condition only in the middle of the nanowire at $x=0$ and not at the nanowire ends at $x=\pm L$. This limit is valid  if the length $2L$ of the nanowire length is much larger than the  localization lengths 
of the FBSs and MBSs. As a result, only spatially decaying eigenstates of $\mathcal H$ can be normalized, thus, we neglect all growing eigenstates.

The four eigenstates decaying in the left segment $x<0$ are given in the basis $\Psi=(\Psi_1, \Psi_{\bar 1},\Psi_1^\dagger, \Psi_{\bar 1}^\dagger)$ by
\begin{widetext}
\begin{align}
&\Phi_{1}^L = \begin{pmatrix}
 i e^{- i (\phi_1+2k_{so,\bar 1}x)} \\ e^{2ik_{so,\bar 1}x}\\- i e^{- i (\phi_1-2k_{so,\bar 1}x)} \\e^{-2ik_{so,\bar 1}x}
\end{pmatrix} e^{\frac{x}{\chi_1}}, \
\Phi_{2}^L = \begin{pmatrix}
 e^{- i (\phi_1+2k_{so,\bar 1}x)} \\ i e^{2ik_{so,\bar 1}x} \\e^{- i (\phi_1-2k_{so,\bar 1}x)} \\-i e^{-2ik_{so,\bar 1}x}
\end{pmatrix} e^{\frac{x}{\chi_1}}, \
\Phi_{3}^L = \begin{pmatrix}
1\\-i e^{-i \phi_{2-}}\\ 1 \\ i e^{-i \phi_{2-}}
\end{pmatrix} e^{\frac{x}{\chi_{2-}}},\ \Phi_{4}^L = \begin{pmatrix}
 -i \\  e^{-i \phi_{2+}}\\ i \\e^{-i \phi_{2+}}
\end{pmatrix}  e^{\frac{x}{\chi_{2+}}},
\end{align}
\end{widetext}
where we used the notations
\begin{align}
&e^{i\phi_1} = (\sqrt{\epsilon^2 - \Delta_{sc}^2} + i \epsilon)/\Delta_{sc},\\
&e^{i\phi_{2\pm}} = (\sqrt{\epsilon^2 - (\Delta_{sc}\pm\Delta_Z)^2} + i \epsilon)/(\Delta_{sc}\pm\Delta_Z).
\end{align}
The localization lengths are given by 
\begin{align}
&\chi_1 = \hbar\upsilon_{F,\bar1}/\sqrt{\epsilon^2 - \Delta_{sc}^2},\\
&\chi_{2\pm} = \hbar\upsilon_{F,\bar1}/\sqrt{\epsilon^2 - (\Delta_{sc}\pm\Delta_Z)^2}. 
\end{align}

\begin{figure}[!bt]
\includegraphics[width=0.7\linewidth]{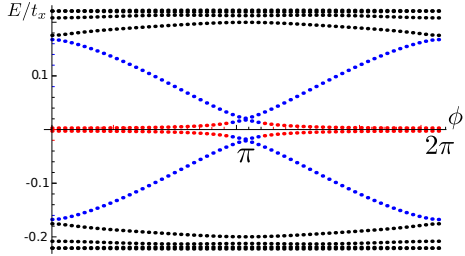}
\caption{The same as in Fig. \ref{no_overlap} except that the chain is shorter with $N=30$.
If the FBSs localized at $x=0$ (blue dots) are away from zero energy, the MBSs localized at $x=\pm L$ (red dots) stay at zero energy. However, if the FBS energy gets close to zero, the two MBSs hybridize with each other via the FBS states, and, thus, split away from zero energy.
}
\label{split}
\end{figure}

Similarly, we can find the four eigenstates decaying in the right segment of the nanowire, {\it i.e.}, for $x>0$,
\begin{widetext}
\begin{align}
&\Phi_{1}^R = \begin{pmatrix}
 e^{ -2 i k_{so, 1}x} \cos (\phi/2)+e^{ -i (\phi_1-2k_{so, 1}x)} \sin (\phi/2) \\ 
i e^{-2ik_{so,1} x} \sin (\phi/2)  -i  e^{ -i (\phi_1-2k_{so, 1}x)} \cos (\phi/2) \\
e^{2ik_{so, 1}x} \cos(\phi/2)+  e^{  -i (\phi_1+2ik_{so, 1})x} \sin (\phi/2) \\
 -i e^{2ik_{so,1} x} \sin (\phi/2)  +i  e^{ -i (\phi_1+2k_{so, 1}x)} \cos (\phi/2)
\end{pmatrix} e^{-\frac{x}{\chi_1}},\ \Phi_{3}^R = \begin{pmatrix}
\cos(\phi/2) -e^{i \phi_{2-}}\sin (\phi/2)\\ 
 i [e^{i \phi_{2-}}\cos (\phi/2) +\sin(\phi/2)]\\
 \cos(\phi/2)-e^{i \phi_{2-}}\sin (\phi/2) \\
- i [e^{i \phi_{2-}}\cos (\phi/2) +\sin(\phi/2)]
\end{pmatrix} e^{-\frac{x}{\chi_{2-}}}
 \nonumber\\
 & \Phi_{2}^R = \begin{pmatrix}
-i e^{ -2 i k_{so, 1}x} \cos (\phi/2) + i  e^{ -i (\phi_1 -2 k_{so, 1}x)}\sin (\phi/2) \\ 
e^{ -2 i k_{so, 1}x} \sin (\phi/2) + e^{ -i (\phi_1 -2 k_{so, 1}x)}\cos (\phi/2)\\
i e^{ 2 i k_{so, 1}x} \cos (\phi/2) - i  e^{ -i (\phi_1 +2 k_{so, 1}x)}\sin (\phi/2) \\ 
e^{ 2 i k_{so, 1}x} \sin (\phi/2) + e^{ -i (\phi_1 + 2 k_{so, 1}x)}\cos (\phi/2)
\end{pmatrix} e^{-\frac{x}{\chi_1}}, \
\Phi_{4}^R = \begin{pmatrix}
 - i [\cos(\phi/2) + e^{i \phi_{2+}}\sin (\phi/2)]\\ 
-e^{i \phi_{2+}}\cos (\phi/2) +\sin(\phi/2)\\
 i[\cos(\phi/2)+e^{i \phi_{2+}}\sin (\phi/2)] \\
 -e^{i \phi_{2+}}\cos (\phi/2) +\sin(\phi/2)
\end{pmatrix} e^{-\frac{x}{\chi_{2+}}},
\end{align}
\end{widetext}
where the Fermi velocity $\upsilon_{F,1}$ should be used instead of $\upsilon_{F,\bar 1}$ in the above expressions for the localization lengths.

Next, we impose the boundary condition on the combination of these eigenstates. If the boundary conditions can be satisfied at some energy $\epsilon$, then there is a bound state. In general, the FBS wavefunction should be a linear superposition of decaying eigenstates,
\begin{align}
\Phi(x)=\begin{cases}
\sum_i a_i \Phi_{i}^L, & x\leq0,\\
\sum_i b_i \Phi_{i}^R, & x\geq0,
\end{cases}
\end{align}
where $a_i$ and $b_i$ are determined by the boundary conditions. The boundary condition on the wavefunction at $x=0$ are given by the continuity condition
\begin{align}
&\Phi_\sigma(x=0^+)=\Phi_\sigma(x=0^-),\\
&\Phi_\sigma^\dagger(x=0^+)=\Phi_\sigma^\dagger(x=0^-).
\end{align}
In addition, by integrating the Schroedinger equation for the total Hamiltonian [see Eq. (\ref{sum})] around $x=0$, we find the condition on the derivatives of the wavefunction,
\begin{align}
&\partial_x\Phi_\sigma(x=0^+)-\partial_x\Phi_\sigma(x=0^-)= \nonumber\\
&\hspace{30pt} i (m/\hbar^2 )  [(\boldsymbol{\alpha}_1  -\boldsymbol{\alpha}_{\bar 1}  )\cdot \boldsymbol{\sigma}]_{\sigma\sigma'} \Phi_{\sigma'}(x=0),\\
&\partial_x\Phi_\sigma^\dagger (x=0^+)-\partial_x\Phi_\sigma^\dagger(x=0^-)= \nonumber\\
&\hspace{30pt}- i(m/\hbar^2 )  [(\boldsymbol{\alpha}_1  -\boldsymbol{\alpha}_{\bar 1}  )\cdot \boldsymbol{\sigma}^T]_{\sigma\sigma'} \Phi_{\sigma'}^\dagger(x=0),
\end{align}
with implied summation over repeated spin indices.
The discontinuity in the derivatives arises from the discontinuity of the SOI term at $x=0$. 

In what follows, we assume that the nanowire is in the topological regime such that $\Delta_z>\Delta_{sc}$.  Two MBSs are localized at the nanowire ends: one at the left  end $x=-L$ and one at the right  end $x=L$, and both are not affected by the change in the SOI vector rotation vector $\phi$, see Fig. \ref{closing}. These MBSs stay at zero energy as long as they do not overlap with each other\cite{diego_numerics,DasSarma} and with possible FBSs localized around $x=0$ (compare Fig. \ref{no_overlap} and Fig. \ref{split}). In this work we focus on FBSs and for details on MBS wavefunctions we refer to Ref.~\onlinecite{MF_composite}.

To begin with, we show that if the SOI vectors are opposite in the two sections, {\it i.e.}$, \phi=\pi$, the interface at $x=0$ hosts two zero-energy bound states,  {\it i.e.}, two MBSs.~\cite{pi_junction} The corresponding wavefunctions $\Phi_{MBS}^{(i=1,2)}(x)$ are of the form, written in the basis $(\Psi_1, \Psi_{\bar1},\Psi_1^\dagger, \Psi_{\bar1}^\dagger)$,
\begin{align}
\Phi_{MBS}^{(i)}(x)=\begin{pmatrix}
f_i(x)\\
g_i(x)\\
f_i^{*}(x)\\
g_i^{*}(x)
\end{pmatrix}.
\end{align}
If we also assume for the sake of simplicity that $\alpha_1=\alpha_{\bar 1}=\alpha$ in what follows, then the functions $f_i$ and $g_i$ are given by
\begin{align}
&f_1= \begin{cases}
e^{-2ik_{so}x} e^{x/\xi_1} - e^{x/\xi_{2-}},& x<0\\
e^{-x/\xi_{2-}} -e^{2ik_{so}x} e^{-x/\xi_1}, &x>0
\end{cases},\nonumber\\
&g_1 = i f_1^*,
\label{strong_wave}
\end{align}
\begin{widetext}
\begin{align}
&f_2=\begin{cases}
 e^{-2ik_{so}x } e^{x/\xi_1}  (1+ i k_{so} \xi_1) (k_{so} \xi_1+k_{so} \xi_{2-}) - 2e^{x/\xi_{2-}} [1 +  ( k_{so} \xi_1)^2 ](k_{so} \xi_{2-}), & x<0\\
 i e^{2ik_{so} x} e^{-x/\xi_1}(1+ i k_{so} \xi_1) [2 ( k_{so} \xi_1) ( k_{so} \xi_{2-}) -i ( k_{so} \xi_1- k_{so} \xi_{2- }) ],& x>0
\end{cases}\nonumber,\\
&g_2 = i f_2^*,
\label{MF2}
\end{align}
\end{widetext}
where we have introduced the SOI wavevector $k_{so}=m \alpha/\hbar^2$.
As can be seen from Eq. (\ref{MF2}), the wavefunctions are involved even for the simple special case of $\phi=\pi$.

\begin{figure}[!b]
\includegraphics[width=0.7\linewidth]{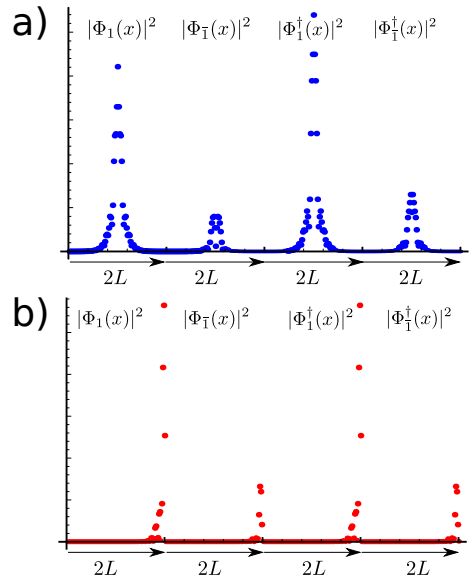}
\caption{The wavefunction components of the FBS [panel (a), blue dots] and of the MBS [panel (b), red dots]  in the basis $(\Psi_1,\Psi_{\bar1},\Psi_1^\dagger,\Psi_{\bar1}^\dagger)$ obtained numerically from the tight-binding model. The SOI vector is rotated by the angle $\phi=\pi/2$ in the right section. All other parameters are the same as in Fig. \ref{no_overlap}.}
\label{WF_MF}
\end{figure}

Next, we focus on the deviations of the rotation angle $\phi$ from the special value $\phi=\pi$. The accidental two-fold degeneracy 
for FBSs being filled and unfilled  gets lifted, and the corresponding energy level goes away from zero. In general, the eigenvalue equation obtained from the boundary conditions is too involved to be solved analytically for all parameter values (see Fig. \ref{WF_MF}). However, one can demonstrate that the energy of the FBS  grows linearly as the SOI rotation angle slightly deviates from $\phi=\pi$,
\begin{align}
\epsilon(\phi) = \pm \Delta_{sc} \left(1-\frac{\Delta_{sc}}{\Delta_Z}\right) (\phi-\pi).
\end{align}
A general remark is in order here.
The energies of the FBSs are {\it excitation} energies, and thus only the non-negative energy values are physical, whereas the negative energies are spurious resulting from
the formal doubling of the degrees of freedom in the Nambu representation used here. Still, following standard practice, we will always show the full spectrum with positive and negative energies.

In addition, in the special case $\Delta_Z=2\Delta_{sc}$, the problem simplifies considerably such that the FBS spectrum can be found analytically for all angle values and is given by
\begin{align}
\epsilon = \pm \Delta_{sc} \cos (\phi/2).
\label{ep_sc}
\end{align}
In general, the FBS energy moves inside the energy gap as a function of $\phi$. When the SOI vectors are aligned in  opposite directions, $\phi=\pi$, the bound states are at zero energy. However, if the SOI vectors are not collinear, the FBS energy is finite, see Fig. \ref{no_overlap}. Moreover, the FBS disappear in the continuum if $\phi=0$ when there is no interface around $x=0$ and the SOI is (locally) uniform.

\section{Regime of Weak Spin-Orbit Interaction \label{sec:weak}}

\subsection{Linearized Effective Hamiltonian}

In this section, we focus on the weak SOI regime (or alternatively, on the regime of a strong magnetic field) in which the Zeeman energy dominates over the SOI energy and the superconducting pairing, $\Delta_Z \gg E_{SOI},\ \Delta_{sc}$. In this case, the spins are almost aligned along the magnetic field in the $x$-direction. The SOI vector,
which lies in the $yz$-plane at the angle $\phi$ to the $z$ axis, tilts the spins slightly into the $yz$ plane. As a result, the wavefunction at the Fermi level in the absence of the superconductivity is given in the basis $(\Psi_1, \Psi_{\bar 1})$ by 
\begin{align}
\phi_0^{R/L}=\frac{1}{\sqrt{2}}\begin{pmatrix}
                              -1\pm e^{i\phi}\frac{k_{so}}{k_F}\\
                              1\pm e^{i\phi}\frac{k_{so}}{k_F}
                             \end{pmatrix} 
                            e^{\pm i k_F x},\label{rt}
\end{align}
where we only keep terms up to linear order in $k_{so}/{k_F}$. Here, the Fermi wavevector $\pm k_F$ is solely determined 
by the magnetic field, $k_F=\sqrt{2m\Delta_Z}/\hbar$. The plus (minus) sign corresponds to the right (left) mover with the wavevector $k_F$ ($-k_F$). 

The kinetic energy is written in terms of slowly-varying right [$R(x)$] and left [$L(x)$] movers as 
\begin{align}
H_{kin}^w= -i\hbar \upsilon_F \int dx \ [R(x)^\dagger \partial_x R(x)-	L(x)^\dagger \partial_x L(x)].
\label{weak_kin}
\end{align}

The superconducting term is obtained directly from Eq.~(\ref{SC}),
\begin{align}
H_{SC}^w= \int dx\ \bar\Delta_{sc} [ e^{-i\phi} R^\dagger(x) L^\dagger(x) +H.c.],
\label{weak_sc}
\end{align}
where the superconductivity strength is given by
\begin{align}
\bar \Delta_{sc} = e^{i\phi} \Delta_{sc} (\phi^R_0)^*\cdot (i\sigma_2) (\phi_0^L)^*= 2   \Delta_{sc} \frac{k_{so}}{k_F}.
\end{align}

Again, the spectrum of the total Hamiltonian $H^w=H_{kin}^w+H_{SC}^w$ is independent of the direction of the SOI vector in the $yz$ plane,
\begin{equation}
E_\pm=\pm\sqrt{(\hbar \upsilon_F k)^2+\bar \Delta_{sc}^2}.
\end{equation}

\subsection{Bound States}

Next, we explore the presence of localized states at the interface at $x=0$ where the SOI vector changes its direction. 
Without loss of generality, we assume that $\phi$ is fixed to zero in the negative section of the nanowire and can vary from $0$ to $2\pi$ in the positive section.
Two decaying eigenstates at energy $\epsilon$  are given by
\begin{align}
\Phi_{\pm L} = \begin{pmatrix}
           \pm \frac{\epsilon\pm i\sqrt{\bar\Delta_{sc}^2-\epsilon^2}}{\bar\Delta_{sc}}\\1
            \end{pmatrix}
e^{\pm ik_F x} e ^{x/\xi_\epsilon}
\label{phiL}
\end{align}
in the left section of the nanowire $x<0$ and by
\begin{align}
\Phi_{\pm R} = \begin{pmatrix}
         \pm   \frac{ \epsilon\mp i\sqrt{\bar\Delta_{sc}^2-\epsilon^2}}{\bar\Delta_{sc}} \\ e^{i\phi}
            \end{pmatrix}
e^{\pm ik_F x} e ^{-x/\xi_\epsilon}.
\label{phiR}
\end{align}
in the right section of the nanowire $x>0$.
Here, the localization length depends on the 
separation of the energy level $\epsilon$ from the bulk gap and is given by $\xi_\epsilon=\hbar\upsilon_F/\sqrt{\bar\Delta_{sc}^2-\epsilon^2}$.

The spectrum of the localized states is given by
\begin{align}
\epsilon_{\pm}(\phi)=\pm \bar \Delta_{sc} \cos (\phi/2).
\label{abrupt_weak_SOI}
\end{align}
This analytical formula is in good agreement with the numerical results shown in Fig. \ref{no_overlap}.
The corresponding wavefunctions are given by
\begin{align}
&\Phi_+(x) = \begin{pmatrix}
             e^{-i\phi/2} \\ 1
            \end{pmatrix} e^{ik_Fx} e^{- |x|/\xi} \label{weak_wave} \\
&\Phi_-(x) =\begin{pmatrix}
              e^{-i\phi/2}\\1
            \end{pmatrix}e^{-ik_Fx} e^{- |x|/\xi}    \, ,           
\end{align}
where the localization length is given by $\xi = \hbar \upsilon_F/ [\bar \Delta_{sc} \sin (\phi/2)]$.
Here, the basis $(\Psi_0,\Psi^\dagger_0)$ is composed of the operator $\Psi_0$ that correspond to the annihilation
operator acting on electrons in the lowest Zeeman-field-split subband that are spin-polarized along the $x$-axis, see Eq.~(\ref{rt}).

We confirm again that the FBSs are at zero-energy if $\phi=\pi$.\cite{pi_junction} Moreover, the FBS energy level moves smoothly as
a function of the SOI rotation angle inside the gap $\bar \Delta_{sc}$ until it disappears in the bulk spectrum
at $\phi=0$. As expected, the localization length $\xi_\epsilon$ is the longer the closer the energy level is to the edge of the bulk gap.
Interestingly, the electron and hole components of the FBS wavefucntions $\Phi_\pm(x)$ differ only by the phase irrespective of the energy
$\epsilon_{\pm}$. Thus, the charge of the FBSs is zero. However, we recall that this result is valid only as long as corrections
of order ${k_{so}}/{k_F}$
are neglected. To check this property, we also calculated the charge numerically and found that the non-zero charge corrections are at most a few percent of $e$ and thus indeed negligibly small.

\begin{figure*}[bt!]
\includegraphics[width=0.85\linewidth]{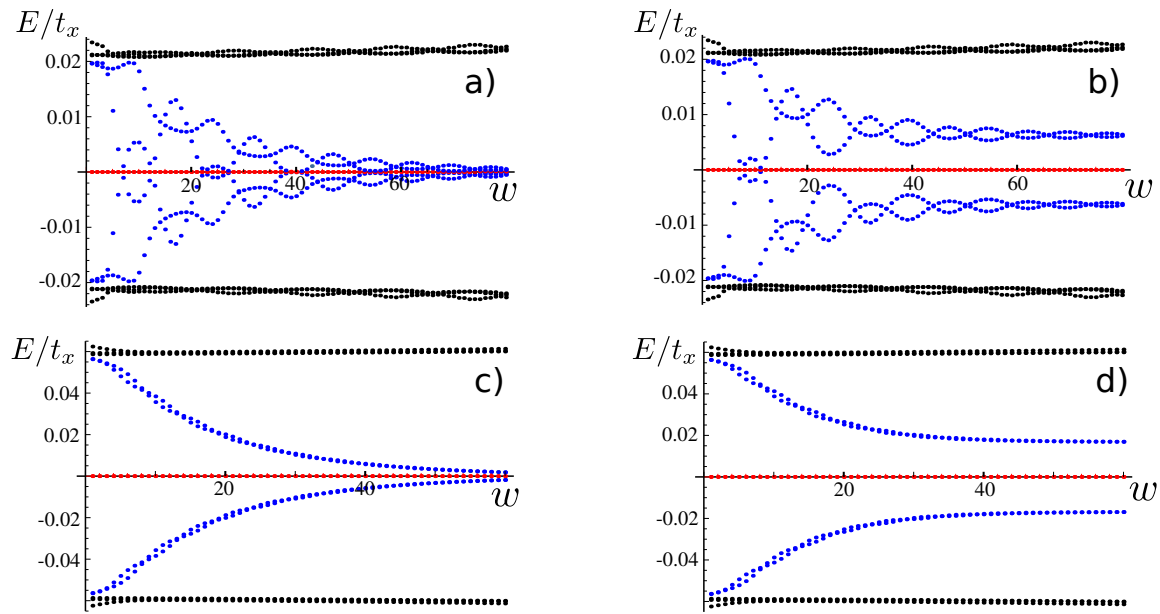}
\caption{The energy spectrum of the nanowire in the topological phase with a double SOI-junction as function of the junction half-width $w$ as numerically obtained in the tight-binding model [see Eqs. (\ref{tb2}) - (\ref{tb1})]. The chain consists of $N=300$ sites. The SOI vector rotation along the chain is described by Eq. (\ref{angle}) with  $\phi_0=\pi$ [panels (a),(c)] and  $\phi_0=4\pi/5$ [panels (b) and (d)].
The two initially independent FBS (blue dots) hybridize as the double junction gets shorter. Panels (a) and (b): Strong SOI regime ($\bar \alpha/t_x=0.2$, $\Delta_{sc}/t_x=0.02$, and $\Delta_Z/t_x=0.05$), the energy of hybridized states exhibits Friedel oscillations and even periodically returns to zero-energy.
Panels (c) and (d): Weak SOI regime ($\bar \alpha/t_x=0.1$, $\Delta_{sc}/t_x=0.2$, and $\Delta_Z/t_x=0.4$), Friedel oscillations are nearly visible, and the FBS level  smoothly merges with the bulk (black dots)
as $w$ gets small. The MBSs (red dots) at each wire end coexist with the FBSs.
}
\label{junction}
\end{figure*}

\section{Double SOI-Junction \label{sec:junction}}
In this section, we consider a double junction composed of two SOI-junction considered above.
The two junctions are separated by a distance $2w$ from each other such that the SOI vector is given by
\begin{align}
\boldsymbol{\alpha}(x)= \begin{cases}
          \alpha_{\bar 1} \hat {\bf z}, & -L<x<-w\\
          \alpha_{1} (\bf \hat z \cos \phi   +\bf \hat y \sin  \phi  ), &-w<x<w\\
          \alpha_{\bar 1} \hat {\bf z},  & w<x<L\\
         \end{cases}.
         \label{angle}
\end{align}
Here, we assume the double junction to be symmetric around $x=0$ such that the FBSs are at the same 
energy at each of two SOI-junctions in the limit of large $w$, {\it i.e.}, $w\gg \xi$. First, such a situation maximizes the degree of hybridization between them
(compare with the hybridization between FBSs and MBSs in Fig. \ref{split}).
Second, this allows us to go the limit of small separation, $w \approx \lambda_F$, where the effect of the SOI-junctions effectively vanishes.
In this limit, the nanowire becomes again spatially uniform, and thus the FBSs will merge with the bulk states, see Fig.~\ref{junction}. 
On the other hand, in the intermediate regime of $w\approx \xi$, the two overlapping FBSs hybridize into  symmetric and antisymmetric states
with energies distinct from the ones obtained for a single SOI-junction. The energy splitting between the two hybridized states depends on the
system parameters as shown in Fig. \ref{junction}.

In the regime of strong SOI, the components of the FBS wavefunction [see, for example, Eq. (\ref{strong_wave})] not only exponentially  decay
away from the SOI-junction but also oscillate with the wavevector $2k_F$. As a consequence, the overlap between the two FBS wavefunctions, and thus the splitting
between them, is also an oscillating function, see Figs. \ref{junction}a and \ref{junction}b. Moreover, as expected,
these Friedel oscillations are highly sensitive to the  width $w$ and get stronger as the double junction shrinks. The amplitude of the oscillations can become so
large that the FBS energy could even go back to zero. 

In the opposite regime of weak SOI, however,
the structure of the  FBS wavefunction [see Eq. (\ref{weak_wave})] is such that Friedel oscillations are nearly absent.
This is in good agreement with the numerical results presented in Figs. \ref{junction}c and \ref{junction}d.
The two hybridized FBS have almost the same energy and smoothly merge with the bulk states with vanishing width $w$.

\section{Smooth rotation of SOI \label{sec:smooth}}

In contrast to the previous sections, in which
the SOI vector changes its direction abruptly at $x=0$, in this section,
the SOI vector rotates smoothly as function of position $x$.
Without loss of generality,  we assume that the SOI vector rotation 
angle $\phi(x)$ is given by
\begin{equation}
\phi (x)= \begin{cases}
  0, & x\leq0,\\
  \phi_0 x/\ell, & 0<x<\ell,\\
  \phi_0, & x\geq \ell,
 \end{cases}
\end{equation}
where $\ell$ is a characteristic length that determines the adiabaticity of the SOI vector rotation.
The choice of the linear change of the phase as a function of the coordinate $x$ is motivated by the fact that in this particular case we can
find explicit eigenstates of the effective Hamiltonian $H_{kin}^w+H_{sc}^w$ in the regime of weak SOI [see Eqs. (\ref{weak_kin}) and (\ref{weak_sc})].
Two decaying eigenstates in the left and right sections are given by Eq. (\ref{phiL}) and Eq. (\ref{phiR}) correspondingly.
In the central section of the nanowire $0<x<\ell$, there are four eigenstates
\begin{align}
\Phi_{\pm C}^{(p)}= \begin{pmatrix}
            e^{-i\phi_0 x/2\ell} \\ e^{i\phi_0 x/2\ell \mp i \theta_p} 
           \end{pmatrix} e^{ip k_Fx } e^{\mp x/\xi_{\epsilon,p}},
\end{align}
 where the localization lengths are given by
\begin{align}
\xi_{\epsilon,p}=\frac{2L}{\sqrt{(2\bar\Delta_{sc}\ell/\hbar \upsilon_F)^2-(2\epsilon\ell/\hbar \upsilon_F+p\phi_0)^2}}.
\end{align}
Here, the label $p$ takes the values $\pm1$. We also introduced the new notation
\begin{widetext}
\begin{align}
&e^{\pm i\theta_p}=\frac{2E\ell/\hbar \upsilon_F+p\phi_0\pm i \sqrt{(2\bar\Delta_{sc}\ell/\hbar \upsilon_F)^2-(2\epsilon\ell/\hbar \upsilon_F+p\phi_0)^2}}{2\bar\Delta_{sc}\ell/\hbar \upsilon_F}.
\end{align}
\end{widetext}

\begin{figure}[!tb]
\includegraphics[width=0.78\linewidth]{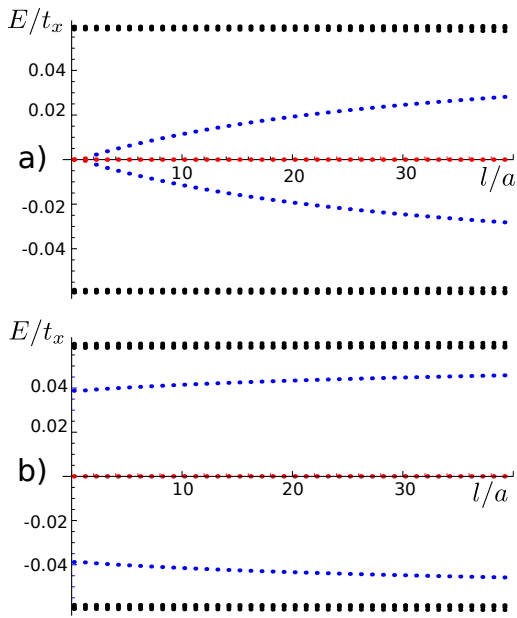}
\caption{The energy spectrum of the nanowire in the topological phase as function of the characteristic  SOI width $l$ as numerically found in the tight-binding model [see Eqs. (\ref{tb2}) - (\ref{tb1})]. The chain consists of $N=300$ sites. The parameters are fixed to be $\bar \alpha/t_x=0.1$, $\Delta_{sc}/t_x=0.2$, and $\Delta_Z/t_x=0.4$,
and $\mu=E_{SOI}$. The SOI vector rotates in total by the angle  $\phi=\pi$ [(a)] and  $\phi=\pi/2$ [(b)]. 
The MBSs (red dots)
are not affected by the width of the SOI. In contrast, the FBS energy (blue dots) is sensitive to  $l$. 
The smoother the variation the closer is the FBS energy to the edge of the gap (black dots). 
In general, if the SOI variation is smooth on the Fermi wavelength scale $\lambda_F$ (here, $\lambda_F/a\approx 10$), the FBS  level merges with the bulk.}
\label{pi2}
\end{figure}

After finding the eigenstates of the Hamiltonian, we follow  standard procedure
and impose the boundary conditions on the wavefunction $\Phi(x)$ and its derivative at the two interfaces
\begin{align}
&\Phi (x=0^+)=\Phi(x=0^-),\label{bo_1}\\
&\partial_x\Phi (x=0^+)=\partial_x\Phi(x=0^-),\\
&\Phi (x=\ell^+)=\Phi(x=\ell^-),\\
&\partial_x\Phi(x=\ell^+)=\partial_x\Phi(x=\ell^-)\label{bo_2},
\end{align}
where $\Phi(x)$ is defined as a linear combination
\begin{align}
\Phi(x) = \begin{cases}
\sum_{j=\pm 1} a_j \Phi_{j R} & x\leq 0,\\
\sum_{j=\pm 1} (c_j \Phi_{j C}^{(1)}+ d_j \Phi_{j C}^{(2)}) & 0<x<\ell,\\
\sum_{j=\pm 1} b_j \Phi_{j L} & x\geq L\, .
\end{cases}
\end{align}
If Eqs. (\ref{bo_1}) - (\ref{bo_2}) can be satisfied at some energy $\epsilon$, $\epsilon<\Delta_{sc}$, for non-zero values for the coefficients $a_j$, $b_j$, $c_j$, and $d_j$, 
then there is a bound state of the energy $\epsilon$ localized at the interface $x\in(0,\ell)$. The general expression for the bound state energy is too involved to be reproduced here, and, instead, we
show only a limiting case.

We consider a limit in which the SOI vector rotates over distances much smaller than the Fermi wavelength $\lambda_F=2\pi/k_F$, $\ell\ll \lambda_F$, and, thus, much smaller than
the localization length $\xi_\epsilon$.
This limit is the closest to the case of an abrupt change considered above [see Eq. (\ref{abrupt_weak_SOI})]. We find that there is a trend to the repulsion of the energy levels from the zero energy towards to the bulk gap, see Fig.~\ref{pi2}. In particular, the twofold  degeneracy of zero-energy states at $\phi=\pi$ gets lifted. The energy spectrum is given by
\begin{equation}
E_{\pm} = \pm \bar\Delta_{sc} \cos \left(\frac{\phi}{2}\right)  \pm \frac{\bar\Delta_{sc}^2 \ell}{\hbar \upsilon_F}\frac{2-2\cos\phi-\phi \sin \phi}{2 \phi}.
\end{equation}
All these finding are  consistent with our numerical results, see Fig.~\ref{pi2}. The energy level disappears in the bulk 
if the SOI vector rotates more slowly than any  characteristic length in the problem, $\ell>\xi_E\gg1/k_F$.

\section{Conclusions and Outlook \label{sec:conc}}

We have studied  FBSs in  nanowires with a spatially non-uniform SOI and found that they can coexist with MBSs in the topological phase.
In particular,
we explored the dependence of their energy on the rotation angle $\phi$ of the SOI vector along the nanowire. In general, the FBS state can assume any energy value inside the gap depending on $\phi$, in contrast to MBSs which are truly stable and remain at zero energy as long as the MBSs do not overlap.
Specifically, if the SOI vector rotates abruptly, the FBS energy level moves from zero at $\phi=\pi$ to the gap energy at $\phi=0$.
However, if the SOI rotates adiabatically the FBSs are absent (merged with the bulk states). 
Non-uniform SOI with FBSs can arise from varying Rashba SOI induced by varying electric field directions along the wire,~\cite{Rashba_1960,SOI_book,Fasth,Rashba_SOI,Kloeffel_Trif}
 from Dresselhaus SOI~\cite{Dresselhaus_1958,SOI_book} due to discontinuities in the crystallographic structure 
(like in $T$-junctions of wires), or from domain walls in rotating Zeeman fields~\cite{Rotating_field,pi_junction} or RKKY systems~\cite{RKKY_Basel,RKKY_Simon,RKKY_Franz} 
proposed for MBSs. 
In our work we focussed on solid-state systems as mentioned above. However, we note that the effects discussed here can also be implemented in cold atom and optical lattice systems~\cite{cold_atoms}
where  topological phases in one-dimensional systems of the type considered here have been proposed.~\cite{demler_2011}

Next, under certain conditions, FBSs can interfere with  topological quantum computing schemes which have been proposed for MBSs in wire networks.~\cite{Alicea_Fisher}
As an outlook for further work, let us mention briefly three such scenarios in a qualitative manner.

First, it was shown that non-overlapping MBSs can hybridize via the medium, {\it e.g} via the bulk superconductor which induces the proximity gap in the nanowire. \cite{Sasha}
Similarly, FBSs could also serve as an effective mediator of hybridization between two distant MBSs which can have wavefunction overlap via the fermionic intragap state. However, this coupling is efficient
only for FBSs at energies close to zero (see Fig. \ref{split}) since otherwise the overlap of wavefunctions is exponentially suppressed.

Second, the FBSs can trap quasiparticles and contribute to the decoherence time of logical qubits based on MBSs.\cite{Diego} At any finite temperature,  quasiparticle excitations in the bulk superconductor \cite{quasiparticles_1,quasiparticles_2,quasiparticles_3}
tend to minimize their energy  by occupying lower energy states such as the ones, for instance, provided by the
FBS described in this work.
Consider now a  pair of well-separated MBSs that is filled with a quasiparticle. For braiding one envisions to move the MBSs through the wire.~\cite{Alicea_Fisher}
In particular, if a MBS now is moved through a nanowire section that hosts a  FBS filled with a trapped quasiparticle, then there is a finite probability that the quasiparticle shared by the MBS pair and the quasiparticle localized in the FBS overlap and form a Cooper pair which then is absorbed by the superconducting condensate. 
As a result, the parity of the MBS subsystem changes and thus the coherence of the logical qubit based on MBSs is lost.

Third, if the FBS sits at zero energy, it can be represented as two MBSs. If one additional MBS is passing through this FBS (during a braiding operation), it might not be guarantueed in general that these two MBS forming the FBS will return to their initial state at the end of
the operation, with the third MBS again being far away from the FBS.
If this is not the case, then one might expect that the braiding will be affected non-trivially by the presence of such a FBS. This then might be in contrast to the situation when all (three) MBS are moved together.~\cite{akhmerov} However, this question requires further investigation which is beyond the scope of this work.

Next, we also considered symmetric double junctions and showed that they contain double FBS that can hybridize. This behavior is very much reminiscent of double quantum dots,~\cite{Hanson,Kloeffel}
which suggests that such double FBS, being effectively spinless, can serve as a promising platform for conventional charge qubits. Indeed, as we have seen, the amount of charge of an FBS is rather small, 
especially in the weak SOI regime. This implies that the direct coupling to electric field fluctuations due to {\it e.g.}  external voltage gates will  be correspondingly small. 
On the other hand, fluctuations of a Rashba SOI field will also induce level fluctuations of the FBS and thus act as dephasing source. This might be less the case for double junctions induced by non-uniform Dresselhaus SOI which is independent of gates. Nevertheless, these considerations suggests that such charge qubits, embedded in a topological superconductor, might be rather
well protected against environmental noise and, thus, enjoy unusually large dephasing times, especially the closer their energy is to zero. 

Finally, the presence and location of the FBS can serve as a detection tool of possible discontinuities of the SOI, whereby a transport measurement of the type discussed in Refs.~\onlinecite{FF_transport,FF_pump} can reveal the energy of the FBS and thus the rotation angle of the SOI-junction.

To conclude, the understanding of FBSs is not only interesting on its own, but such bound states can also have a variety of interesting effects and applications which deserve further exploration. It also seems worthwhile to search for them experimentally.

\section*{Acknowledgments}
We acknowledge support from the Harvard Quantum Optical Center and from the Swiss NSF and NCCR QSIT.


\begin{thebibliography}{100}





\bibitem{fu} L. Fu and C. L. Kane, Phys. Rev. Lett. {\bf 100}, 096407 (2008).

\bibitem{Nagaosa_2009} Y. Tanaka, T. Yokoyama, and N. Nagaosa, Phys. Rev. Lett. {\bf 103}, 107002 (2009).

\bibitem{Sato}M. Sato and S. Fujimoto, Phys. Rev. B {\bf 79}, 094504 (2009).

\bibitem{lutchyn_majorana_wire_2010} R. M. Lutchyn, J. D. Sau, and S. Das Sarma,
Phys. Rev. Lett. {\bf 105}, 077001 (2010).

\bibitem{oreg_majorana_wire_2010} Y. Oreg, G. Refael, and F. von Oppen,
Phys. Rev. Lett. {\bf 105}, 177002 (2010).

\bibitem{alicea_majoranas_2010} J. Alicea, Phys. Rev. B {\bf 81}, 125318 (2010).

\bibitem{MF_ee_Suhas} S. Gangadharaiah, B. Braunecker, P. Simon, and D. Loss, Phys. Rev. Lett. {\bf 107}, 036801 (2011).

\bibitem{potter_majoranas_2011} A. C. Potter and P. A. Lee, Phys. Rev. B {\bf 83}, 094525 (2011).

\bibitem{Klinovaja_CNT} J. Klinovaja, S. Gangadharaiah, and D. Loss, Phys. Rev. Lett. {\bf 108}, 196804 (2012).

\bibitem{Pascal} D. Chevallier, D. Sticlet, P. Simon, and C. Bena, Phys. Rev. B {\bf 85}, 235307 (2012).

\bibitem{bilayer_MF_2012} J. Klinovaja, G. J. Ferreira, and D. Loss, Phys. Rev. B  {\bf 86}, 235416 (2012).

\bibitem{Bena_MF} D. Sticlet, C. Bena, and P. Simon, Phys. Rev. Lett. {\bf 108}, 096802 (2012).

\bibitem{Rotating_field} J. Klinovaja, P. Stano, and D. Loss, Phys. Rev. Lett. {\bf 109}, 236801 (2012).

\bibitem{Ali} S. Nadj-Perge, I. K. Drozdov, B. A. Bernevig, and A. Yazdani, Phys. Rev. B {\bf 88}, 020407(R) (2013).

\bibitem{RKKY_Basel} J. Klinovaja, P. Stano, A. Yazdani, and D. Loss, Phys. Rev. Lett. {\bf 111}, 186805 (2013).

\bibitem{RKKY_Simon} B. Braunecker and P. Simon, Phys. Rev. Lett. {\bf 111}, 147202 (2013).

\bibitem{RKKY_Franz} M. Vazifeh and M. Franz,  Phys. Rev. Lett. {\bf 111}, 206802 (2013).

\bibitem{MF_nanoribbon} J. Klinovaja and D. Loss, Phys. Rev. X {\bf 3}, 011008 (2013).

 \bibitem{MF_Bena} C. Dutreix, M. Guigou, D. Chevallier, and C. Bena, arXiv:1309.1143.

\bibitem{MF_MOS} J. Klinovaja and D. Loss, Phys. Rev. B {\bf 88}, 075404 (2013).

\bibitem{Shiba_ladder} K. Poyhonen, A. Weststrom, J. Rontynen, and T. Ojanen,  Phys. Rev. B {\bf 89}, 115109 (2014).

\bibitem{MF_dot} E. Vernek, P.H. Penteado, A. C. Seridonio, and J. C. Egues, Phys. Rev. B {\bf 89} 165340 (2014).

\bibitem{mourik_signatures_2012}
V. Mourik, K. Zuo, S. M. Frolov, S. R. Plissard, E. P. A. M. Bakkers, and L. P. Kouwenhoven, Science, {\bf 336}, 1003 (2012). 

\bibitem{deng_observation_2012} 
M. T. Deng, C. L. Yu, G. Y. Huang, M. Larsson, P. Caroff, and H. Q. Xu, Nano Lett. {\bf 12}, 6414 (2012).

\bibitem{das_evidence_2012}
A. Das, Y. Ronen, Y. Most, Y. Oreg, M. Heiblum, and H. Shtrikman, Nat. Phys. {\bf 8}, 887 (2012). 

\bibitem{Rokhinson} L. P. Rokhinson, X. Liu, and J. K. Furdyna, Nat. Phys. {\bf 8}, 795 (2012).

\bibitem{Goldhaber} J. R. Williams, A. J. Bestwick, P. Gallagher, S. S.
Hong, Y. Cui, A. S. Bleich, J. G. Analytis, I. R. Fisher,
and D. Goldhaber-Gordon, Phys. Rev. Lett. {\bf 109}, 056803
(2012).

\bibitem{marcus_MF} H. O. H. Churchill, V. Fatemi, K. Grove-Rasmussen, M.
Deng, P. Caroff, H. Q. Xu, and C. M. Marcus, Phys. Rev.
B {\bf 87}, 241401(R) (2013).



\bibitem{JR_model} R. Jackiw and C. Rebbi, Phys. Rev. D {\bf 13}, 3398 (1976).

\bibitem{CDW_suhas} S. Gangadharaiah, L. Trifunovic, and D. Loss, Phys. Rev.
Lett. {\bf 108}, 136803 (2012).

\bibitem{FF_1} R. Rajaraman and J. S. Bell, Phys. Lett. B {\bf 116}, 151
(1982).

\bibitem{SSH_model} W. P. Su, J. R. Schrieffer, and A. J. Heeger, Phys. Rev. Lett. {\bf 42}, 1698 (1979).


\bibitem{FF_non_Abelian} J. Klinovaja and D. Loss, Phys. Rev. Lett. {\bf 110}, 126402
(2013).

\bibitem{FF_transport} D. Rainis, A. Saha, J. Klinovaja, L. Trifunovic, and D. Loss,
Phys. Rev. Lett. {\bf 112}, 196803 (2014). 

\bibitem{FF_pump} A. Saha, D. Rainis, R. Tiwari, and D. Loss, Phys. Rev. B {\bf 90}, 035422 (2014).



\bibitem{Fradkin_PF_1980} E. Fradkin and L. P. Kadanoff,  Nucl.
Phys. B {\bf 170}, 1 (1980).

\bibitem{topology_barkeshli} M. Barkeshli and X.-L. Qi,
Phys. Rev. X {\bf 2}, 031013 (2012).


\bibitem{Fendley_PF_2012} P. Fendley,  J. Stat. Mech. {\bf 2012}, 11020 (2012).

\bibitem{Cheng} M. Cheng, Phys. Rev. B {\bf 86}, 195126 (2012).

\bibitem{PF_Linder} N. Lindner, E. Berg, G. Refael, and A. Stern, Phys.
Rev. X {\bf 2}, 041002 (2012).

\bibitem{barkeshli_2} M. Barkeshli, C, Jian, and X.-L. Qi, Phys. Rev. B {\bf 87}, 045130.

\bibitem{Vaezi} A. Vaezi, Phys. Rev. B {\bf 87}, 035132 (2013).

\bibitem{PF_Clarke} D. Clarke, J. Alicea, and K. Shtengel, Nat. Commun.
{\bf 4}, 1348 (2013).

\bibitem{Ady_FMF} Y. Oreg, E. Sela, and A. Stern, Phys. Rev. B {\bf 89}, 115402 (2014).

\bibitem{PF_Mong} R. S. K. Mong, D. J. Clarke, J. Alicea, N. H. Lindner, P.
Fendley, C. Nayak, Y. Oreg, A. Stern, E. Berg, K. Shtengel,
and M. P. A. Fisher, Phys. Rev. X {\bf 4}, 011036 (2014).

\bibitem{PFs_Loss} J. Klinovaja and D. Loss, Phys. Rev. Lett. {\bf 112}, 246403 (2014).

\bibitem{vaezi_2} A. Vaezi, Phys. Rev. X {\bf 4}, 031009 (2014).

\bibitem{PFs_Loss_2} J. Klinovaja and D. Loss, Phys. Rev. B {\bf 90}, 045118 (2014).

\bibitem{PF_Oreg} E. Sagi and Y. Oreg, arXiv:1403.1791.

\bibitem{PF_TI} J. Klinovaja, A. Yacoby, and D. Loss, arXiv:1403.4125.

\bibitem{Vaezi_2} A. Vaezi and M. Barkeshli, arXiv:1403.3383.

\bibitem{PF_Thomas} C. Orth, R. Tiwari, T. Meng, T. Schmidt, arXiv:1405.4353.


\bibitem{Stern_review} C. Nayak, S. H. Simon, A. Stern, M. Freedman, and
S. Das Sarma, Rev. Mod. Phys. {\bf 80}, 1083 (2008).

\bibitem{Alicea_review} J. Alicea, Rep. Prog. Phys. {\bf 75}, 076501 (2012).

\bibitem{pi_junction} T. Ojanen, Phys. Rev. B {\bf 87}, 100506 (2013).


\bibitem{Rashba_1960} E. I. Rashba, Sov. Phys. Solid State 2, 1109 (1960).

\bibitem{SOI_book} R. Winkler, {\it Spin-Orbit Coupling Effects in Two-Dimensional Electron and Hole Systems}
(Springer, Berlin, 2003).


\bibitem{Dresselhaus_1958} G. Dresselhaus, Phys. Rev. 100, 580 (1955).



\bibitem{Alicea_Fisher} J. Alicea, Y. Oreg, G. Refael, F. von Oppen, and
M. P. A. Fisher,  Nat. Phys. {\bf 7}, 412 (2011).

\bibitem{Braunecker} B. Braunecker, G. I. Japaridze, J. Klinovaja, and D. Loss,
Phys. Rev. B {\bf 82}, 045127 (2010).

\bibitem{MF_composite}  J. Klinovaja and D. Loss, Phys. Rev. B \textbf{86}, 085408 (2012).

\bibitem{diego_numerics} D. Rainis, L. Trifunovic, J. Klinovaja, and D. Loss, Phys.
Rev. B {\bf 87}, 024515 (2013).

\bibitem{DasSarma} S. Das Sarma, J. D. Sau, and T. D. Stanescu, Phys. Rev. B {\bf 86}, 220506.

\bibitem{Fasth} C. Fasth, A. Fuhrer, L. Samuelson, V. N. Golovach, and
D. Loss, Phys. Rev. Lett. {\bf 98}, 266801 (2007).

\bibitem{Rashba_SOI} 	J. Klinovaja, M. J. Schmidt, B. Braunecker, and D. Loss, Phys. Rev. Lett. {\bf 106}, 156809 (2011).

\bibitem{Kloeffel_Trif} C. Kloeffel, M. Trif, and D. Loss, Phys. Rev. B {\bf 84}, 195314 (2011).


\bibitem{cold_atoms} M. Lewenstein, A. Sanpera, V. Ahufinger, B. Damski, A. Sen, and U. Sen, Adv. Phys. {\bf 56}, 243 (2007).

\bibitem{demler_2011} L. Jiang, T. Kitagawa, J. Alicea, A. Akhmerov, D. Pekker, G. Refael, J. I. Cirac, E. Demler, 
M. D. Lukin, and P. Zoller, Phys. Rev. Lett. {\bf 106}, 220402 (2011).

\bibitem{Sasha} A.A. Zyuzin, D. Rainis, J. Klinovaja, and D. Loss, Phys. Rev. Lett. {\bf 111}, 056802 (2013).


\bibitem{Diego} D. Rainis and D. Loss, Phys. Rev. B {\bf 85}, 174533 (2012).

\bibitem{quasiparticles_1} P. J. de Visser, J. J. A. Baselmans, P. Diener, S. J. C. Yates, A. Endo, and T. M. Klapwijk, Phys. Rev. Lett. {\bf 106}, 167004 (2011).

\bibitem{quasiparticles_2} J. Aumentado, M. W. Keller, J. M. Martinis, and M. H. Devoret, Phys.Rev.Lett. {\bf 92}, 066802 (2004).

\bibitem{quasiparticles_3} L. Sun, L. DiCarlo, M. D. Reed, G. Catelani, L. S. Bishop, D. I. Schuster, 
B. R. Johnson, G. A. Yang, L. Frunzio, L. I. Glazman, M. H. Devoret, and R. J. Schoelkopf, Phys. Rev. Lett. {\bf 108}, 230509 (2012).


\bibitem{akhmerov} A. R. Akhmerov, Phys. Rev. B {\bf 82}, 020509(R).

\bibitem{Hanson} R. Hanson, L. P. Kouwenhoven, J. R. Petta, S. Tarucha, and L. M. K. Vandersypen,  Rev. Mod. Phys. {\bf 79}, 1217 (2007).

\bibitem{Kloeffel} C. Kloeffel and D. Loss, Annu. Rev. Condens. Matter Phys. {\bf 4}, 51 (2013).

\end{thebibliography}
\end{document}